\documentclass[aps,twocolumn,groupedaddress,showpacs]{revtex4}
\usepackage{graphicx}

 \newcommand{\slrr}      {$T_1^{-1}$}
 \newcommand{\cecoin}    {CeCoIn$_5$}
 \newcommand{\ceirin}    {CeIrIn$_5$}
 \newcommand{\cemin}     {CeMIn$_5$}
 \newcommand{\cerhin}     {CeRhIn$_5$}

 \newcommand{\spc}       {superconductor}
 \newcommand{\afc}       {antiferromagnetic}
 \newcommand{\af}        {antiferromagnet}
 \newcommand{\pmc}       {paramagnetic}
 
 \newcommand{\Hint}       {$H_{\rm int}$}

\begin{document}
\bibliographystyle{apsrev}
\preprint{LA-UR-02-1265}

\thispagestyle{myheadings} \markright{{\em LA-UR-02-1265}}

\title{Low Frequency Spin Dynamics in the \cemin\ Materials}
\author{N. J. Curro}
\author{J. L. Sarrao}
\author{J. D. Thompson}
\affiliation{Condensed Matter and Thermal Physics, Los Alamos
National Laboratory, Los Alamos, NM 87545, USA}
\author{P. G. Pagliuso}
\affiliation{Instituto de F\'{\i}sica ``Gleb Wataghin'', UNICAMP,
13083-970, Campinas-SP, Brazil}
\author{\v S. Kos}
\author{Ar. Abanov}
\author{D. Pines}
\affiliation{Theoretical Division, Los Alamos National Laboratory,
Los Alamos, NM 87545, USA}

\date{\today}

\begin{abstract}

We measure the spin lattice relaxation of the In(1) nuclei in the
\cemin\ materials, extract quantitative information about the low
energy spin dynamics of the lattice of Ce moments in both \cerhin\
and \cecoin, and identify a crossover in the normal state.  Above
a temperature $T^*$ the Ce lattice exhibits "Kondo gas" behavior
characterized by local fluctuations of independently screened
moments; below $T^*$ both systems exhibit a "Kondo liquid" regime
in which interactions between the local moments contribute to the
spin dynamics.  Both the \afc\ and superconducting ground states
in these systems emerge from the "Kondo liquid" regime. Our
analysis provides strong evidence for quantum criticality in
\cecoin.

\end{abstract}


\maketitle

The \cemin\ heavy fermion materials possess a rich phase diagram
revealing a fascinating interplay between the \afc\ behavior of
incompletely screened Ce local moments, an unconventional normal
state, and superconducting behavior of the heavy electrons that is
reminiscent of that found in the cuprates
\cite{helmut,PetrovicCo}. The non-Fermi liquid behavior of the
itinerant heavy electrons and the possible d-wave symmetry of
their superconducting state has led to the proposal that the
superconducting  pairing mechanism arises from their coupling to
spin fluctuations
\cite{helmut,lonzarich,spinflucreview,sasha,VarmaSpinFluc}. In
this Letter, we consider the information about the coupled system
of Ce local moments and itinerant heavy electrons that can be
derived from Nuclear Magnetic Resonance (NMR) experiments on the
spin lattice relaxation rate (\slrr) of the In(1) nuclei located
in the plane of the Ce moments.  We show that the In(1) nuclei are
strongly coupled to their four nearest neighbor Ce spins by an
anisotropic hyperfine interaction, and that the resulting
anomalous behavior of $T_1T$ provides important information on the
low frequency dynamics of the lattice of Ce spins, and the
quasiparticles to which they couple. Because the coupling is
anisotropic, it does not vanish for antiferromagnetically
correlated Ce moments, so that \slrr\ provides valuable
information on the dynamics of their magnetic ordering, as well as
on the influence of  Kondo screening of the moments on their
relaxation rate, and departures from Kondo behavior at low
temperatures \cite{baoCeRhIn5v3,coxchi}. We present our results
for both the \af\ \cerhin\ and the \spc\ \cecoin.

Previous reports of \slrr\ in \ceirin\ and \cecoin\ by other
authors  have attributed an unusual temperature dependence of
\slrr\ in the normal state to antiferromagnetic fluctuations up to
room temperature \cite{kitaokaIr115,koharaCeCoIn5}. We show in
both \cerhin\ and \cecoin\ that above a temperature $T^*$ of the
order 10K, \slrr\ is dominated by {\it local} fluctuations of the
Ce moments. Below $T^*$ the {\bf q}-independent local moment
contribution to \slrr\ becomes temperature independent, and and a
second {\bf q} and $T$ dependent component emerges.  We identify
this emergent component with the heavy electrons, that is the
itinerant component of the Ce 4$f$ electrons arising from their
coupling to one another and to the conduction electrons. This
extra contribution to the relaxation agrees quantitatively with
inelastsic neutron scattering (INS) results in \cerhin, and
suggests that the antiferromagnetic correlations in  \cecoin\ are
primarily 2D, and that in the absence of superconductivity one
would have a quantum critical point corresponding to a transition
from Fermi liquid to antiferromagnetic behavior at $T=0$.

In order to develop a model for the hyperfine coupling in the
\cemin\ materials, it is necessary to determine the number of spin
degrees of freedom, and the hyperfine couplings to each degree of
freedom. For all three materials (M=Rh, Ir, Co), the bulk
susceptibility $\chi$ is dominated at high temperatures by
localized Ce moments, and can be adequately fit by an expression
for these moments in a tetragonal crystalline electric field (CEF)
\cite{onukiCEF,curroCeCoIn5,pagliusoCEF,saturo,LawrenceCEF}.
Neutron diffraction (ND) in the ordered state of \cerhin\
indicates that the magnetism is localized on the Ce sites, albeit
with reduced moments \cite{baoCeRhIn5}. The reduced moment
plausibly reflects some degree of hybridization with incomplete
screening of the Ce moment in the CEF ground state doublet. (For
the CEF scheme measured in \cite{LawrenceCEF} the ground state
doublet has a moment of 0.92$\mu_B$, whereas the measured moment
is 0.74$\mu_B$ \cite{baoCeRhIn5,baoCeRhIn5v2,Anna}.) The magnetic
shifts of the high symmetry In(1) site in \cerhin, presented in
Fig. (1), are linear in $\chi$ in the \pmc\ state.  If the main
contribution to the magnetic shift arises from the localized Ce
moments, then
$K_{\alpha}=K_{0,\alpha}+\beta_{\alpha}\chi_{\alpha}$, where
$\beta_{\alpha}$ is an effective hyperfine coupling and
$\alpha=c,ab$. Note that the hyperfine coupling is anisotropic
($\beta_{c}=26.4$kOe/$\mu_B$, $\beta_{ab}=19.6 $kOe/$\mu_B$), and
the size of these hyperfine couplings are more than an order of
magnitude greater than typical dipolar fields ($\sim$ 1
kOe/$\mu_B$), so the In(1) must be coupled to the Ce via an
anisotropic transferred hyperfine coupling of electronic origin.

\begin{figure}
  \centering
  \includegraphics[width=\linewidth]{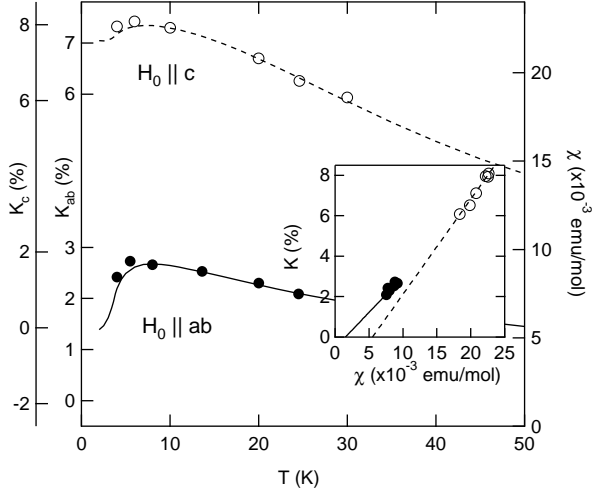}
  \caption{The magnetic shift of In(1) in \cerhin\ (circles) and the
  bulk susceptibility (lines) versus temperature.
  The inset shows $K_{\alpha}$ versus
   $\chi_{\alpha}$. The solid (open) circles and solid (dashed)
  lines are for $H||ab$ ($H||c$).}
  \label{fig:expt}
  \end{figure}

 We postulate that the In(1)
nucleus is coupled to each of the four nearest neighbor Ce spins
through a tensor $\mathbf{\tilde{A}}= \left\{ A_{||}, A_{\perp},
A_c \right\}$, where the principal axes lie parallel and
perpendicular to the Ce-In(1) bond axis in the $ab$ plane, and
along the $c$ axis:
\begin{equation}
\mathcal{H}= \gamma\hbar\sum_{i\in{\rm n.n.}
}\mathbf{I}\cdot\mathbf{\tilde{A}}_i\cdot\mathbf{S}_i.
\end{equation}
Here $\gamma$ is the gyromagnetic ratio, ${\mathbf I}$ is the
nuclear spin of the In(1), ${\mathbf S}_i$ is the spin on the
$i^{th}$ Ce, and the sum is over the four nearest neighbor Ce
sites. Note that there should also exist a direct hyperfine
coupling between the In nuclei and the conduction electrons.
However, the fact that \slrr\ drops by almost two orders of
magnitude below $T_N$ suggests that this coupling is small
compared to the coupling between the nuclei and the Ce moments.
For notational simplicity, we define $A_{ab}$ and $x$ as
$A_{||,\perp}=A_{ab}(1\pm x)$. Using Eq. (1) one can show that
$K_c= 4A_c\chi_c$ and $K_{ab}= 4A_{ab}\chi_{ab}$; using the slopes
of $K_{\alpha}$ versus $\chi_{\alpha}$ given above, we have
$A_{ab} = 4.9 {\rm kOe}/\mu_{\rm B}$ and $A_c = 6.6 {\rm
kOe}/\mu_{\rm B}$. The parameter $x$ can be determined from the
internal field in the \afc\ state. The moments are localized on
the Ce sites, with the structure given by: ${\mathbf S}({\mathbf
r}_i) =S_0 \cos(\pi x/a)\cos(\pi
y/a)\left\{\cos(q_0z),\sin(q_0z),0\right\}$, where
$S_0=0.74\mu_B$, and $q_0=0.297\frac{2\pi}{c}$
\cite{CurroNJ:Rh115magstruc,baoCeRhIn5,baoCeRhIn5v2,Anna}. For
this structure, the internal field is given by: $\mathbf{H}_{int}=
2A_{ab}xS_0\left\{\sin(q_0z),\cos(q_0z),0\right\}$. The measured
internal field \Hint\ at the In(1) site for $T<<T_N$ is 1.7 kOe
\cite{CurroNJ:Rh115magstruc}, which corresponds to a value
$x=0.12$.

\begin{figure}
  \centering
  \includegraphics[width=\linewidth]{T1fig.eps}
  \caption{(a) $T_1T$ versus $T$ in \cerhin\ and \cecoin; the \cerhin\ data for $T>20$K
  are taken from \cite{koharaCeRhIn5} (\cerhin\ NQR ($\bullet$),
  \cecoin\ $H_0||c$ (solid squares), and $H_0||ab$
  (open squares)).
  (b) $(T_1T)^{-1} - 111.43{\rm sec}^{-1}{\rm K}^{-1}$ ($\bullet$) and $\xi/a$ (open squares,
  ref \cite{baoCeRhIn5v3})
  vs. $T-T_N$ in \cerhin, where
  the solid line is a fit as described
  in the text. (c) $\Gamma$ vs. $T$ for \cerhin. The dashed line
  is a linear fit to the data above 20K,
  and the solid line is taken from \cite{coxchi} using $T_0=5$K.}
  \end{figure}

The spin lattice relaxation of the In(1) nuclei is determined by
the fluctuations of the Ce spins, and is given by the Fourier
component of $\langle{\mathcal H}(t){\mathcal H}(0)\rangle$ at the
Larmor frequency, where the time dependence arises from the
fluctuations of the Ce spins ${\mathbf S}_i$.  Moriya showed that
the spin lattice relaxation rate can be expressed in the general
form \cite{moriya}:
\begin{equation}
\frac{1}{T_1T} = \frac{\gamma^2k_B}{2}\lim_{\omega\rightarrow 0}
\sum_{\mathbf{q},\alpha}F_{\alpha}^2(\mathbf{q})
\frac{\chi_{ab}"(\mathbf{q},\omega)}{\omega}
\end{equation}
where $\alpha$ is summed over the two directions perpendicular to
the applied static field.  Here $F_{\alpha}^2({\mathbf q})$ are
form factors, $\chi_{\alpha}(\mathbf{q},\omega)$ is the dynamical
susceptibility, and the sum is over the Brillouin zone. The form
factors are the spatial Fourier transforms of the coupling in Eq.
(1), and are given by:
\begin{eqnarray}
F_{ab}^2(\mathbf{q})&=&16A_{ab}^2\cos^2\left(\frac{q_xa}{2}\right)\cos^2\left(\frac{q_ya}{2}\right)\\
\nonumber&& + 16A_{ab}^2x^2\sin^2\left(\frac{q_xa}{2}\right)\sin^2\left(\frac{q_ya}{2}\right)\\
F_c^2(\mathbf{q})&=&16A_c^2\cos^2\left(\frac{q_xa}{2}\right)\cos^2\left(\frac{q_ya}{2}\right).
\end{eqnarray}
Since $F_{ab}^2(\mathbf{q})>0$ for all $\mathbf{q}$, \slrr\ will
pick up fluctuations at all wavevectors.  In particular, the In(1)
is sensitive to the critical slowing down of the spin fluctuations
above $T_N$.

We now can determine quantitatively the low frequency spin
spectrum of \cerhin. As may be seen in Fig. (2), $T_1T$ varies
quadratically with $T$ for $T>20$K, which is what might have been
anticipated for a collection of independent Ce moments that are
weakly coupled to the conduction electrons via an interaction
${\mathcal H}=J{\mathbf \sigma}\cdot{\mathbf S}$. To see this we
note that the susceptibility of the local moments is given by:
\begin{equation}
\chi_L(\omega)=\frac{\chi_0(T)}{1-i\omega/\Gamma(T)},
\end{equation}
where $\chi_0(T)$ is the bulk susceptibility ($\sim 1/T$ for
$T>20$K), $\Gamma(T)=\pi J^2N^2(0)k_BT$, and $N(0)$ is the
electronic density of states at $E_F$. From Eqs. (2,3,5) we then
have:
\begin{equation}
(T_1T)^{-1} = 4\gamma^2 k_BA_{ab}^2(1+x^2)\chi_0(T)/\Gamma(T) \sim
T^{-2}.
\end{equation}
Fig. (2c) shows $\Gamma(T)$  as a function of temperature. Note
that $\Gamma\sim T$ for $T>20$K, with a slope determined by
$JN(0)\sim0.24$. Although INS data of $\Gamma(T)$ is unavailable
in this material, Bao and coworkers have estimated that
$\Gamma<<3$meV for $T< 7$K \cite{baoCeRhIn5v3}.

We interpret the crossover at 20K to an almost constant value of
$\Gamma$ as reflecting the onset of Kondo screening of the local
moments. The relaxation of independently screened local moments, a
"Kondo gas",  has been calculated by Cox {\it et al}. \cite{cox},
who find $\Gamma(T)=T_0f(T/T_0)$ where $f$ is a universal function
reflecting a temperature dependent effective coupling, and $T_0$
is proportional to the Kondo temperature. For $T>T_0$, Qachaou
{\it et al}. have shown that this scaling behavior holds for
several mixed-valent and heavy electron materials \cite{qachaou}.
As may be seen in Figs. (2c) and (3), a good fit to the
experimental data may be obtained with $T_0=5$K over the range
8K$<T<$40K.

Below $T^* \sim$ 8K one begins to see the "Kondo liquid" expected
when the interaction between the Ce local moments induced by their
coupling to the conduction electrons plays a dominant role in the
relaxation of the moments. Indeed, for $T_N<T<T^*$, \slrr\ shows a
strong divergence associated with the critical slowing down of the
spin fluctuations of this Kondo liquid. In this region the
susceptibility takes the form:
\begin{equation}
\chi_{AF}({\bf q}, \omega)= {\alpha \xi ^2(T) \mu_B^2 \over 1 +
\xi ^2(T)({\bf q-Q})^2-{i\omega/ \omega _{sf}(T)}},
\end{equation}
which adequately describes the INS data \cite{baoCeRhIn5v3}. Here
$\alpha$ is a temperature independent constant, and
$\omega_{sf}\sim 1/\xi^z$, where $z$ is the dynamical scaling
constant. In the mean field regime, $z=2$ and we have
$(T_1T)^{-1}= (T_1T)^{-1}_0 + (T_1T)^{-1}_{AF}$, where
$(T_1T)^{-1}_0$ is the $\mathbf{q}-$independent, "local",
contribution assumed constant below $T^*$, and $(T_1T)^{-1}_{AF}$
is the contribution from the \afc\ fluctuations.  The latter is
given by $(T_1T)^{-1}_{AF} \approx \gamma^2 A_{ab}^2x^2
k_B\alpha\mu_B^2/2\pi^2\omega_{sf}\xi$, where we have assumed 3D
fluctuations, and that the correlation length along the $c$
direction, $\xi_c\sim\xi_{ab}\sim\xi$. If we assume mean field
behavior, then $(T_1T)^{-1}_{AF}\sim \xi(T)$. Fig. (2b) shows
$(T_1T)^{-1} -(T_1T)^{-1}_0$ versus $T$, where $(T_1T)_0 = {\rm
\;}0.00897$ sec K. The solid line is a fit to $\xi\sim
(T-T_N)^{-\nu}$, where $\nu=0.30\pm0.05$. As seen in Fig. (2b),
the temperature dependence of the NMR measurements agrees well
with INS measurements of $\xi$. Note that for a single critical
point, a mean field analysis gives $\nu=\beta=1/2$ ($M_{\rm
sublattice} \sim (T_N-T)^{\beta}$). The fact that we observe
$\nu\sim 0.3$ suggests that either the point $T_N(H=0)$ is a
multicritical point, or that fluctuations change the exponents.
The former is a plausible explanation for the critical exponent
$\beta=0.25$ observed in NMR and ND measurements
\cite{CurroNJ:Rh115magstruc,baoCeRhIn5}, and is consistent with
the phase diagram in \cite{Rh115phasediagram}.

\begin{figure}
  \centering
  \includegraphics[width=\linewidth]{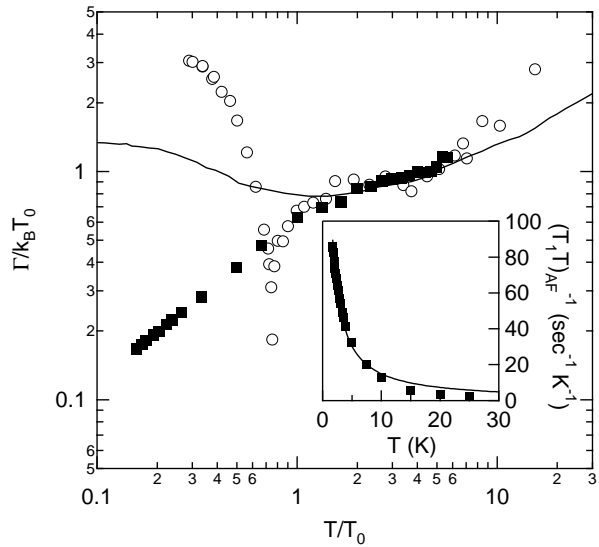}
  \caption{$\Gamma(T)/k_BT_0$ versus $T/T_0$.  ($\circ$)
  \cerhin\ with $T_0=5$K, and (solid squares) \cecoin\ with $T_0=17$K. The
  solid line is from Cox \cite{coxchi}. The inset shows the build-up of antiferromagnetic
  correlations below $\sim$30K follows the $1/T$ behavior (solid line) described in the text, and
  extend to $T_c=1.8$K, the transition temperature for the 5T magnetic field used in this experiment. }
  \end{figure}

We next consider the superconductor \cecoin.  In order to deduce
the hyperfine tensor, three independent experimental quantities
are required. From the Knight shifts in this material measured in
\cite{curroCeCoIn5} we have $A_{ab} = 3.0 {\rm kOe}/\mu_{\rm B}$
and $A_c = 2.2 {\rm kOe}/\mu_{\rm B}$.  As discussed there, the
Knight shift measurements show that for $T<50K$ $A_c$ vanishes
whereas $A_{ab}$ does not change; this behavior is reflected as
the anomalous rise in the $T_1T$ data below 40K in Fig. (2a) for
$H_0||ab$.  A direct measurement of the parameter $x$ is lacking
since there is no magnetic order \cite{comment}. For concreteness
we assume $x=0.12$, as in \cerhin. Fig. (3) shows
$\Gamma(T)/k_BT_0$ versus $T/T_0$ for \cecoin\ and \cerhin, where
$T_0=5$K for \cerhin, and $T_0=17$K, for \cecoin. The \cecoin\
data were determined using the $T_1T$ data for $H_0||c$. For
different values of $x$, the overall magnitude of the $\Gamma(T)$
data, and hence $T_0$, will change by at most a factor of two, but
the overall temperature dependence will be the same and the
conclusions will not change.

Note that in both materials the data scale with the Cox formula
for $T>T^* \approx 1.5T_0$, however below this temperature
$\Gamma(T)$ is less than the Cox prediction. As was the case for
\cerhin, $\Gamma(T)$ drops markedly below the Cox "Kondo gas"
prediction below $T^*$, a change we can identify with the onset of
a "Kondo liquid" regime in which the interactions between the
local moments gives rise to a new physical regime. We suggest that
just as in the Rh material, $T^*$ in the Co material also marks
the onset of antiferromagnetic correlations, and proceed to
analyze the data in \cecoin\ below $T^*$ in the same fashion as in
\cerhin.

The inset of Fig. (3) shows $(T_1T)_{AF}^{-1}=(T_1T)^{-1} -
(T_1T)^{-1}_{0}$ versus $T$, where we have taken
$(T_1T)_{0}=0.0793$ sec K, the value at 30K.  The solid line is
$(T-T_N)^{-1}$, with $T_N=0$. Importantly, this result suggests
that in the absence of superconductivity \cecoin\ would order
magnetically at $T\sim 0$, so that in this material one has a
quantum critical point that is obscured by superconductivity
\cite{nicklas}. Quantum criticality is also consistent with the
non-Fermi liquid behavior seen in the normal state
\cite{romanQCP}. It is also consistent with experiments at fields
$H>H_{c2}$ that show the specific heat increases at low
temperatures as though there were an ordering temperature at $T
\approx 0$ \cite{PetrovicCo}. The temperature dependence of
$(T_1T)_{AF}^{-1}$ is significant. In the \cerhin,
$(T_1T)_{AF,3D}^{-1}\sim\xi$, as expected for 3D fluctuations. On
the other hand, for quasi-2D fluctuations
$(T_1T)_{AF,2D}^{-1}\sim\omega_{sf}^{-1}\sim\xi^2$. In fact, the
data in \cecoin\ are best understood in a picture of 2D
fluctuations, where $\omega_{sf}\sim T$, which is exactly the
result observed in the cuprates above the pseudogap temperature
\cite{curro1248}.  In light of the behavior observed in \cerhin\,
such correlations provide a natural explanation of the suppression
of $\Gamma$, as well as a possible mechanism for the d-wave
superconductivity \cite{monthoux}. Indirect evidence for such
correlations has been seen in \cite{saturo}. Measurements of
Knight shift and \slrr\ under pressure and as a function of doping
should prove invaluable to understanding the evolution of these
antiferromagnetic correlations as the ground state changes from
antiferromagnetic to superconducting.

We thank P. C. Hammel, C. P. Slichter, W. Bao, A. Balatsky, C.
Varma, J. Lawrence, Z. Fisk, and Y. Bang for useful discussions.
This work was performed under the auspices of the U.S. Department
of Energy.


\begin{thebibliography}{29}
\expandafter\ifx\csname
natexlab\endcsname\relax\def\natexlab#1{#1}\fi
\expandafter\ifx\csname bibnamefont\endcsname\relax
  \def\bibnamefont#1{#1}\fi
\expandafter\ifx\csname bibfnamefont\endcsname\relax
  \def\bibfnamefont#1{#1}\fi
\expandafter\ifx\csname citenamefont\endcsname\relax
  \def\citenamefont#1{#1}\fi
\expandafter\ifx\csname url\endcsname\relax
  \def\url#1{\texttt{#1}}\fi
\expandafter\ifx\csname
urlprefix\endcsname\relax\def\urlprefix{URL }\fi
\providecommand{\bibinfo}[2]{#2}
\providecommand{\eprint}[2][]{\url{#2}}

\bibitem[{\citenamefont{Hegger et~al.}(2000)\citenamefont{Hegger, Petrovic,
  Moshopoulou, Hundley, Sarrao, Fisk, and Thompson}}]{helmut}
\bibinfo{author}{\bibfnamefont{H.}~\bibnamefont{Hegger}},
  \bibinfo{author}{\bibfnamefont{C.}~\bibnamefont{Petrovic}},
  \bibinfo{author}{\bibfnamefont{E.~G.} \bibnamefont{Moshopoulou}},
  \bibinfo{author}{\bibfnamefont{M.~F.} \bibnamefont{Hundley}},
  \bibinfo{author}{\bibfnamefont{J.~L.} \bibnamefont{Sarrao}},
  \bibinfo{author}{\bibfnamefont{Z.}~\bibnamefont{Fisk}}, \bibnamefont{and}
  \bibinfo{author}{\bibfnamefont{J.~D.} \bibnamefont{Thompson}},
  \bibinfo{journal}{Phys. Rev. Lett.} \textbf{\bibinfo{volume}{84}},
  \bibinfo{pages}{4986} (\bibinfo{year}{2000}).

\bibitem[{\citenamefont{Petrovic et~al.}(2001)\citenamefont{Petrovic, Pagliuso,
  Hundley, Movshovich, Sarrao, Thompson, Fisk, and Monthoux}}]{PetrovicCo}
\bibinfo{author}{\bibfnamefont{C.}~\bibnamefont{Petrovic}},
  \bibinfo{author}{\bibfnamefont{P.~G.} \bibnamefont{Pagliuso}},
  \bibinfo{author}{\bibfnamefont{M.~F.} \bibnamefont{Hundley}},
  \bibinfo{author}{\bibfnamefont{R.}~\bibnamefont{Movshovich}},
  \bibinfo{author}{\bibfnamefont{J.~L.} \bibnamefont{Sarrao}},
  \bibinfo{author}{\bibfnamefont{J.~D.} \bibnamefont{Thompson}},
  \bibinfo{author}{\bibfnamefont{Z.}~\bibnamefont{Fisk}}, \bibnamefont{and}
  \bibinfo{author}{\bibfnamefont{P.}~\bibnamefont{Monthoux}},
  \bibinfo{journal}{Journal Of Physics-Condensed Matter}
  \textbf{\bibinfo{volume}{13}}, \bibinfo{pages}{L337} (\bibinfo{year}{2001}).

\bibitem[{\citenamefont{Mathur et~al.}(1998)\citenamefont{Mathur, Grosche,
  Julian, Walker, Freye, Haselwimmer, and Lonzarich}}]{lonzarich}
\bibinfo{author}{\bibfnamefont{N.~D.} \bibnamefont{Mathur}},
  \bibinfo{author}{\bibfnamefont{F.~M.} \bibnamefont{Grosche}},
  \bibinfo{author}{\bibfnamefont{S.~R.} \bibnamefont{Julian}},
  \bibinfo{author}{\bibfnamefont{I.~R.} \bibnamefont{Walker}},
  \bibinfo{author}{\bibfnamefont{D.~M.} \bibnamefont{Freye}},
  \bibinfo{author}{\bibfnamefont{R.~K.~W.} \bibnamefont{Haselwimmer}},
  \bibnamefont{and} \bibinfo{author}{\bibfnamefont{G.~G.}
  \bibnamefont{Lonzarich}}, \bibinfo{journal}{Nature}
  \textbf{\bibinfo{volume}{394}}, \bibinfo{pages}{39} (\bibinfo{year}{1998}).

\bibitem[{\citenamefont{Chubukov et~al.}()\citenamefont{Chubukov, Pines, and
  Schmalian}}]{spinflucreview}
\bibinfo{author}{\bibfnamefont{A.~V.} \bibnamefont{Chubukov}},
  \bibinfo{author}{\bibfnamefont{D.}~\bibnamefont{Pines}}, \bibnamefont{and}
  \bibinfo{author}{\bibfnamefont{J.}~\bibnamefont{Schmalian}},
  \bibinfo{note}{cond-mat/0201140}.

\bibitem[{\citenamefont{Bang et~al.}(2002)\citenamefont{Bang, Martin, and
  Balatsky}}]{sasha}
\bibinfo{author}{\bibfnamefont{Y.}~\bibnamefont{Bang}},
  \bibinfo{author}{\bibfnamefont{I.}~\bibnamefont{Martin}}, \bibnamefont{and}
  \bibinfo{author}{\bibfnamefont{A.~V.} \bibnamefont{Balatsky}},
  \bibinfo{journal}{Phys. Rev. B}
  \textbf{\bibinfo{volume}{66}} \bibinfo{pages}{224502}
  (\bibinfo{year}{2002}).

\bibitem[{\citenamefont{Miyake et~al.}(1986)\citenamefont{Miyake, Schmitt-Rink,
  and Varma}}]{VarmaSpinFluc}
\bibinfo{author}{\bibfnamefont{K.}~\bibnamefont{Miyake}},
  \bibinfo{author}{\bibfnamefont{S.}~\bibnamefont{Schmitt-Rink}},
  \bibnamefont{and} \bibinfo{author}{\bibfnamefont{C.}~\bibnamefont{Varma}},
  \bibinfo{journal}{Phys. Rev. B}
  \textbf{\bibinfo{volume}{34}}(\bibinfo{number}{9}), \bibinfo{pages}{6554 }
  (\bibinfo{year}{1986}).

\bibitem[{\citenamefont{Bao et~al.}(2002)\citenamefont{Bao, Aeppli, Lynn,
  Pagliuso, Sarrao, Hundley, Thompson, and Fisk}}]{baoCeRhIn5v3}
\bibinfo{author}{\bibfnamefont{W.}~\bibnamefont{Bao}},
  \bibinfo{author}{et al.},
  \bibinfo{journal}{Phys. Rev. B} \textbf{\bibinfo{volume}{65}},
  \bibinfo{pages}{100505} (\bibinfo{year}{2002}).

\bibitem[{\citenamefont{Cox et~al.}(1985)\citenamefont{Cox, Bickers, and
  Wilkins}}]{coxchi}
\bibinfo{author}{\bibfnamefont{D.}~\bibnamefont{Cox}},
  \bibinfo{author}{\bibfnamefont{N.}~\bibnamefont{Bickers}}, \bibnamefont{and}
  \bibinfo{author}{\bibfnamefont{J.}~\bibnamefont{Wilkins}},
  \bibinfo{journal}{J. Appl. Phys.} \textbf{\bibinfo{volume}{57}},
  \bibinfo{pages}{3166} (\bibinfo{year}{1985}).

\bibitem[{\citenamefont{Zheng et~al.}(2001)\citenamefont{Zheng, Tanabe, Mito,
  Kawasaki, Kitaoka, Aoki, Haga, and Onuki}}]{kitaokaIr115}
\bibinfo{author}{\bibfnamefont{G.}~\bibnamefont{Zheng}},
  \bibinfo{author}{et al.},
  \bibinfo{journal}{Phys. Rev. Lett.} \textbf{\bibinfo{volume}{86}},
  \bibinfo{pages}{4664} (\bibinfo{year}{2001}).

\bibitem[{\citenamefont{Kohori et~al.}(2001)\citenamefont{Kohori, Yamato,
  Iwamoto, Kohara, Bauer, Maple, and Sarrao}}]{koharaCeCoIn5}
\bibinfo{author}{\bibfnamefont{Y.}~\bibnamefont{Kohori}},
  \bibinfo{author}{et al.},
  \bibinfo{journal}{Phys. Rev. B} \textbf{\bibinfo{volume}{64}},
  \bibinfo{pages}{134526} (\bibinfo{year}{2001}).

\bibitem[{\citenamefont{Takeuchi et~al.}(2001)\citenamefont{Takeuchi, Inoue,
  Sugiyama, Aoki, Tokiwa, Haga, Kindo, and Onuki}}]{onukiCEF}
\bibinfo{author}{\bibfnamefont{T.}~\bibnamefont{Takeuchi}},
  \bibinfo{author}{et al.}, \bibinfo{journal}{J.
  Phys. Soc. Jpn.} \textbf{\bibinfo{volume}{70}}, \bibinfo{pages}{877}
  (\bibinfo{year}{2001}).

\bibitem[{\citenamefont{Curro et~al.}(2001)\citenamefont{Curro, Simovic,
  Hammel, Pagliuso, Sarrao, Thompson, and Martins}}]{curroCeCoIn5}
\bibinfo{author}{\bibfnamefont{N.~J.} \bibnamefont{Curro}},
  \bibinfo{author}{et al.}, \bibinfo{journal}{Phys. Rev. B}
  \textbf{\bibinfo{volume}{64}}, \bibinfo{pages}{180514}
  (\bibinfo{year}{2001}).

\bibitem[{\citenamefont{Pagliuso et~al.}(2002)\citenamefont{Pagliuso, Moreno,
  Curro, Thompson, Hundley, Sarrao, and Fisk}}]{pagliusoCEF}
\bibinfo{author}{\bibfnamefont{P.~G.} \bibnamefont{Pagliuso}},
  \bibinfo{author}{\bibfnamefont{N.~O.} \bibnamefont{Moreno}},
  \bibinfo{author}{\bibfnamefont{N.~J.} \bibnamefont{Curro}},
  \bibinfo{author}{\bibfnamefont{J.~D.} \bibnamefont{Thompson}},
  \bibinfo{author}{\bibfnamefont{M.~F.} \bibnamefont{Hundley}},
  \bibinfo{author}{\bibfnamefont{L.~L.} \bibnamefont{Sarrao}},
  \bibnamefont{and} \bibinfo{author}{\bibfnamefont{Z.}~\bibnamefont{Fisk}}
  (\bibinfo{year}{2002}), \bibinfo{note}{to appear in Physica B}.

\bibitem[{\citenamefont{Nakatsuji et~al.}(2002)\citenamefont{Nakatsuji, Yeo,
  Balicas, Fisk, Schlottmann, Pagliuso, Moreno, Sarrao, and Thompson}}]{saturo}
\bibinfo{author}{\bibfnamefont{S.}~\bibnamefont{Nakatsuji}},
  \bibinfo{author}{et al.}, \bibinfo{journal}{Phys. Rev. Lett.}
  \textbf{\bibinfo{volume}{89}}, \bibinfo{pages}{106402}
  (\bibinfo{year}{2002}).

\bibitem[{\citenamefont{Christianson et~al.}(2002)\citenamefont{Christianson,
  Lacerda, Hundley, Pagliuso, and Sarrao}}]{LawrenceCEF}
\bibinfo{author}{\bibfnamefont{A.}~\bibnamefont{Christianson}},
  \bibinfo{author}{\bibfnamefont{A.}~\bibnamefont{Lacerda}},
  \bibinfo{author}{\bibfnamefont{M.}~\bibnamefont{Hundley}},
  \bibinfo{author}{\bibfnamefont{P.}~\bibnamefont{Pagliuso}}, \bibnamefont{and}
  \bibinfo{author}{\bibfnamefont{J.}~\bibnamefont{Sarrao}},
  \bibinfo{journal}{Physical Review B}
  \textbf{\bibinfo{volume}{6605}}(\bibinfo{number}{5}), \bibinfo{pages}{4410}
  (\bibinfo{year}{2002}).

\bibitem[{\citenamefont{Bao et~al.}(2001{\natexlab{a}})\citenamefont{Bao,
  Pagliuso, Sarrao, Thompson, Fisk, Lynn, and Erwin}}]{baoCeRhIn5}
\bibinfo{author}{\bibfnamefont{W.}~\bibnamefont{Bao}},
  \bibinfo{author}{et al.},
  \bibinfo{journal}{Phys. Rev. B} \textbf{\bibinfo{volume}{62}},
  \bibinfo{pages}{14 621} (\bibinfo{year}{2001}{\natexlab{a}}).

\bibitem[{\citenamefont{Bao et~al.}(2001{\natexlab{b}})\citenamefont{Bao,
  Pagliuso, Sarrao, Thompson, Fisk, Lynn, and Erwin}}]{baoCeRhIn5v2}
\bibinfo{author}{\bibfnamefont{W.}~\bibnamefont{Bao}},
  \bibinfo{author}{et al.},
  \bibinfo{journal}{Phys. Rev. B} \textbf{\bibinfo{volume}{63}},
  \bibinfo{pages}{219901} (\bibinfo{year}{2001}{\natexlab{b}}).

\bibitem[{\citenamefont{Llobet}(2002)}]{Anna}
\bibinfo{author}{\bibfnamefont{A.}~\bibnamefont{Llobet}}
  (\bibinfo{year}{2002}), \bibinfo{note}{private communication}.

\bibitem[{\citenamefont{Curro et~al.}(2000)\citenamefont{Curro, Hammel,
  Pagliuso, Sarrao, Thompson, and Fisk}}]{CurroNJ:Rh115magstruc}
\bibinfo{author}{\bibfnamefont{N.~J.} \bibnamefont{Curro}},
  \bibinfo{author}{\bibfnamefont{P.~C.} \bibnamefont{Hammel}},
  \bibinfo{author}{\bibfnamefont{P.~G.} \bibnamefont{Pagliuso}},
  \bibinfo{author}{\bibfnamefont{J.~L.} \bibnamefont{Sarrao}},
  \bibinfo{author}{\bibfnamefont{J.~D.} \bibnamefont{Thompson}},
  \bibnamefont{and} \bibinfo{author}{\bibfnamefont{Z.}~\bibnamefont{Fisk}},
  \bibinfo{journal}{Phys. Rev. B} \textbf{\bibinfo{volume}{62}},
  \bibinfo{pages}{R6100} (\bibinfo{year}{2000}).

\bibitem[{\citenamefont{Kohori et~al.}(2000)\citenamefont{Kohori, Yamato,
  Iwamoto, and Kohara}}]{koharaCeRhIn5}
\bibinfo{author}{et al.},
  \bibinfo{journal}{Eur. Phys. B} \textbf{\bibinfo{volume}{18}},
  \bibinfo{pages}{601} (\bibinfo{year}{2000}).

\bibitem[{\citenamefont{Moriya}(1963)}]{moriya}
\bibinfo{author}{\bibfnamefont{T.}~\bibnamefont{Moriya}}, \bibinfo{journal}{J.
  Phys. Soc. Jpn.} \textbf{\bibinfo{volume}{18}}, \bibinfo{pages}{516}
  (\bibinfo{year}{1963}).

\bibitem[{\citenamefont{Kim et~al.}(1995)\citenamefont{Kim, Makivic, and
  Cox}}]{cox}
\bibinfo{author}{\bibfnamefont{E.}~\bibnamefont{Kim}},
  \bibinfo{author}{\bibfnamefont{M.}~\bibnamefont{Makivic}}, \bibnamefont{and}
  \bibinfo{author}{\bibfnamefont{D.~L.} \bibnamefont{Cox}},
  \bibinfo{journal}{Phys. Rev. Lett.} \textbf{\bibinfo{volume}{75}},
  \bibinfo{pages}{2015 } (\bibinfo{year}{1995}).

\bibitem[{\citenamefont{Qachaou et~al.}(1987)\citenamefont{Qachaou,
  Beaurepaire, Benakki, Lemius, Kappler, Meyer, and Panissod}}]{qachaou}
\bibinfo{author}{\bibfnamefont{A.}~\bibnamefont{Qachaou}},
  \bibinfo{author}{et al.},
  \bibinfo{journal}{J. Mag. Magn. Mat.} \textbf{\bibinfo{volume}{63-64}},
  \bibinfo{pages}{635 } (\bibinfo{year}{1987}).

\bibitem[{\citenamefont{Cornelius et~al.}(2001)\citenamefont{Cornelius,
  Pagliuso, Hundley, and Sarrao}}]{Rh115phasediagram}
\bibinfo{author}{\bibfnamefont{A.}~\bibnamefont{Cornelius}},
  \bibinfo{author}{\bibfnamefont{P.}~\bibnamefont{Pagliuso}},
  \bibinfo{author}{\bibfnamefont{M.}~\bibnamefont{Hundley}}, \bibnamefont{and}
  \bibinfo{author}{\bibfnamefont{J.}~\bibnamefont{Sarrao}},
  \bibinfo{journal}{Phys. Rev. B}
  \textbf{\bibinfo{volume}{6414}}(\bibinfo{number}{14}), \bibinfo{pages}{4411}
  (\bibinfo{year}{2001}).

\bibitem[{\citenamefont{Curro}(2002)}]{comment}
 \bibinfo{note}{In
  principle, one can extract $x$ by measuring the anisotropy of $T_1$ for
  $T>T^*$. Using Eqs. (2-3,6)
  $\frac{(T_1^{-1})_{ab}}{(T_1^{-1})_c}=\frac{1}{2}+\frac{A_c^2}{2A_{ab}^2(1+x%
^2)} \frac{\chi_c(T)}{\chi_{ab}(T)}$. We find this ratio goes to
$1/2$ for
  $T\rightarrow0$, reflecting the fact that $A_c$ vanishes. However, no
  physically reasonable values of $x$ exist that satisfy the data for $T>>50$K,
  which probably signifies that $\Gamma(T)$ is anisotropic.}

\bibitem[{\citenamefont{Sidorov et~al.}(2002)\citenamefont{Sidorov, Nicklas,
  Pagliuso, Sarrao, Bang, Balatsky, and Thompson}}]{nicklas}
\bibinfo{author}{\bibfnamefont{V.}~\bibnamefont{Sidorov}},
  \bibinfo{author}{\bibfnamefont{M.}~\bibnamefont{Nicklas}},
  \bibinfo{author}{\bibfnamefont{P.}~\bibnamefont{Pagliuso}},
  \bibinfo{author}{\bibfnamefont{J.}~\bibnamefont{Sarrao}},
  \bibinfo{author}{\bibfnamefont{Y.}~\bibnamefont{Bang}},
  \bibinfo{author}{\bibfnamefont{A.}~\bibnamefont{Balatsky}}, \bibnamefont{and}
  \bibinfo{author}{\bibfnamefont{J.}~\bibnamefont{Thompson}},
  \bibinfo{journal}{Phys. Rev. Lett.}
  \textbf{\bibinfo{volume}{8915}}(\bibinfo{number}{15}), \bibinfo{pages}{7004}
  (\bibinfo{year}{2002}).

\bibitem[{\citenamefont{Bianchi et~al.}(2002)\citenamefont{Bianchi, Movshovich,
  Oeschler, Gegenwart, Steglich, Thompson, Pagliuso, and Sarrao}}]{romanQCP}
\bibinfo{author}{\bibfnamefont{A.}~\bibnamefont{Bianchi}},
  \bibinfo{author}{et al.},
  \bibinfo{journal}{Phys. Rev. Lett.}
  \textbf{\bibinfo{volume}{8913}}(\bibinfo{number}{13}), \bibinfo{pages}{7002}
  (\bibinfo{year}{2002}).

\bibitem[{\citenamefont{Corey et~al.}(1996)\citenamefont{Corey, Curro, OHara,
  Imai, Slichter, Yoshimura, Katoh, and Kosuge}}]{curro1248}
\bibinfo{author}{\bibfnamefont{R.}~\bibnamefont{Corey}},
  \bibinfo{author}{et al.},
  \bibinfo{journal}{Phys. Rev. B}
  \textbf{\bibinfo{volume}{53}}(\bibinfo{number}{9}), \bibinfo{pages}{5907}
  (\bibinfo{year}{1996}).

\bibitem[{\citenamefont{Monthoux and Lonzarich}(2001)}]{monthoux}
\bibinfo{author}{\bibfnamefont{P.}~\bibnamefont{Monthoux}} \bibnamefont{and}
  \bibinfo{author}{\bibfnamefont{G.~G.} \bibnamefont{Lonzarich}},
  \bibinfo{journal}{Phys. Rev. B} \textbf{\bibinfo{volume}{63}},
  \bibinfo{pages}{054529/1 } (\bibinfo{year}{2001}).

\end{thebibliography}

\end{document}